\documentclass[11pt,a4paper]{article} 
\pdfoutput=1

\usepackage{amsmath,amssymb,amsfonts}
\usepackage{calc}
\usepackage{caption2}
\usepackage{hyperref}
\usepackage[pdftex]{xcolor,graphicx}
\usepackage{cite}
\usepackage{multirow}
\usepackage{ulem}

\numberwithin{equation}{section} 
\setcaptionmargin{1cm}

\long\def\symbolfootnote[#1]#2{\begingroup
\def\thefootnote{\fnsymbol{footnote}}\footnote[#1]{#2}\endgroup}

\graphicspath{{fig/}}
\usepackage{slashed}
\usepackage{amssymb}

\newcommand{\be}{\begin{equation}}
\newcommand{\ee}{\end{equation}}
\newcommand{\bea}{\begin{eqnarray}}
\newcommand{\eea}{\end{eqnarray}}

\newcommand{\rep}[1]{{\bf #1}}

\usepackage{graphics}
\usepackage{mathrsfs}
\usepackage{amssymb}
\usepackage{verbatim}
\usepackage{float}
\usepackage{cancel}

\def\beq{\begin{equation}}
\def\eeq{\end{equation}}
\def\bea{\begin{eqnarray}}
\def\eea{\end{eqnarray}}
\def\bitem{\begin{itemize}}
\def\eitem{\end{itemize}}

\graphicspath{{Fig/}}

\newcommand{\e}{\epsilon}
\newcommand{\re}{r_\epsilon}
\newcommand{\GeV}{\;{\rm GeV}}

\newcommand{\ten}{{\bf 10}}

\newcommand{\refe}[1]{(\ref{#1})}

\begin{document}
\begin{flushright}

\end{flushright}

\vskip 30pt

\begin{center}
{\Large \bf On Composite Two Higgs Doublet Models} \\
\vskip 1.0cm {\large  Enrico Bertuzzo, Tirtha Sankar Ray, Hiroshi de Sandes and Carlos A. Savoy $ {}$} \\[1cm]
{\it ${}$ Institut de Physique Th\'eorique, CEA-Saclay,\\
F-91191 Gif-sur-Yvette Cedex, France. }\\[1.5cm]
\normalsize
\end{center}

\begin{abstract} 

We investigate the issue of anomalous contribution to the T parameter and 
to Flavor Changing Neutral Currents in models with two Higgs doublets 
arising as composite pseudo Nambu-Goldstone modes. The non 
linear Lagrangians of several models are explicitly derived and the anomalous 
contributions to T are identified.  The breaking patterns  $SU(5) 
\rightarrow SU(4)\times U(1)$ and $SU(5)\rightarrow SU(4)$, are analyzed  
first and we show how anomalous contributions to T arise in both 
models.  Apart from that, the embedding of the Standard Model fermions 
in a \rep{10} of $SU(5)$ avoids at the same time large corrections to the 
$Zb\bar{b}$ coupling and Flavor Changing Neutral Current transitions. 
Finally, we propose a model based on the breaking $SO(9)/SO(8)$ that is free 
from anomalous contributions to T and in which the problems of the $Zb\bar{b}$ 
coupling and of Flavor Changing Neutral Currents can be simultaneously solved. 



\end{abstract}


\section{Introduction}

The discovery of a new resonance, with mass of about $125\GeV$ 
and properties compatible with those of a Higgs particle \cite{:2012gk, :2012gu} opens up 
a new era for particle physics, since the exploration of the 
Electroweak Symmetry Breaking (EWSB) sector is just at its 
beginning. A major priority is now the determination of the Higgs 
properties, {\it i.e.} production cross sections and branching ratios.  
This might shed light on the mechanism of EWSB and remove the 
veil off any new physics that may control the tera-scale.

If the Higgs has a non trivial substructure, deviations from the
Standard Model (SM) couplings may be observed, while underlying
symmetries can protect its mass from dangerous quantum
corrections. This simple solution to the gauge hierarchy problem 
is naturally realized if the Higgs fields are pseudo-Goldstone bosons 
(PNGB) arising from a spontaneous breaking $G\rightarrow H$ 
due to some strong dynamics at some scale $f$ \cite{Kaplan:1983fs, 
Georgi:1984ef, Georgi:1984af, Dugan:1984hq}. Once the strong sector 
is integrated out, the residual effective theory can be described in
terms of a non-linear sigma model with the Higgs belonging to the
coset space $G/H$.  Interestingly, some of these models can be 
considered as 4-dimensional duals of 5-dimensional gauge-Higgs 
unified theories realized in $AdS_5$ space. However, the most 
important phenomenology can be extracted from effective
theories with fairly general considerations about the strong sector
and in terms of an expansion in the parameter $\e \equiv v/f$, 
with $v$ the Fermi scale.

Models with minimal \cite{Agashe:2004rs, Contino:2006qr,
Contino:2006nn} and non-minimal \cite{Gripaios:2009pe,
Mrazek:2011iu, Frigerio:2011zg} scalar sectors have been studied in the
literature. Let us notice that several complications are introduced
once non-minimal models are considered.  The most important issue is
to protect the ratio between the vector boson masses.  LEP precision
measurements at the Z pole strongly constraint this ratio, quantified
in term of the oblique electroweak parameter $T$.  Two Higgs
doublet models are notorious for breaking the custodial symmetry that
protect the $T$ parameter. The situation is even worse in composite
models, which are inherently non-renormalizable implying a less
constrained Lagrangian. 

In this paper, we address the $T$ parameter and other issues for
models with two Higgs doublets (2HDM) arising as composite PNGB.
Once the embedding of the SM fermions is defined for each case,
the Glashow-Weinberg prescription \cite{Glashow:1976nt} for natural 
FCNC of the Higgs couplings has to be checked and we exhibit two
models where it is automatically 
realized. Compositeness is also a source of violations of gauge 
coupling universality and the  models are also selected by their ability 
to preserve the (relative) agreement of the SM prediction for the 
$Z\bar{b}_Lb_L$ couplings with experiment. This is ensured
by a rule  on the embedding of the $b$-quark in the models
\cite{Agashe:2006at}. 

The general Lagrangian formalism for the PNGB's was formulated in 
\cite{Coleman:1969sm},  but in the Minimal Composite Higgs Model 
(MCHM)  based on $SO(5)/SO(4)$~\cite{Agashe:2004rs}  the 
non-linear realization of the symmetry has also been conveniently formulated in terms 
of a $SO(5)$ vector 
reminiscent  of a linear realization of the breaking. 
Because of its utility, we generalize here this approach to the 
various cosets related to 2HDM and write in a compact form  the PNGB 
Lagrangian, gauge couplings, Higgs potential and Yukawa couplings 
in terms of the new variables linearly transforming in fundamental 
representations. Then, our analysis of the custodial symmetry and
tree-level contributions to $T$ for the different cosets defining 
2HDM's is simplified by the use of these coordinates. As pointed out
in \cite{Mrazek:2011iu} and as also discussed below,  in the 
$Sp(6)/Sp(4)\times SU(2)$ case with a specific embedding of 
the SM gauge symmetries, there are no contributions to the $T$ 
parameter. Here, we also describe another
$T-$safe model based on the coset $SO(9)/SO(8)$.

For concreteness, we first focus on the symmetry breaking patterns $SU(5)
\rightarrow SU(4)\times H'$ with $H'=U(1)$ or nothing. In the former
case, eight pseudo Goldstone modes are delivered, fitting into two SM 
doublets. In the latter scenario,  the breaking leads to an additional
SM singlet.  We find how the Higgs couplings to $W$ bosons and 
fermions are modified with respect to those of the SM Higgs boson, so
that different production and decay rates are expected. However,
in both cases, only by aligning  the vacuum expectation values 
(vev) of the two Higgs  the contribution to $T$ could vanish.  
The lowest dimensional representations of $SU(5)$, ${\bf 5}$ and
${\bf 10}$, provide a natural embedding for the SM fermions. We
perform a systematic study of all the possibilities, clearly
differentiating between cases that can be successfully embedded into 
a 5d completion and those in which this is not so straightforward.  We find that
the embedding of the SM fermions in the ${\bf 10}$ can lead to a
simultaneous resolution of the usual problems of anomalous
$Z\bar{b}_Lb_L$ \cite{Agashe:2006at} and the flavor problems 
\cite{Agashe:2009di}.  The contribution to the (Coleman-Weinberg) 
PNGB  potential from these fermions is displayed and, in general, 
does not have the symmetry to align the Higgs vev's. 

Finally, we propose a $SO(9)/SO(8)$ model, where the two Higgs 
couplings to the gauge bosons are much like in the MCHM and, 
indeed, there are no tree-level contribution to the $T$ parameter. 
Another interesting feature of this model is an embedding of the SM 
symmetries in $SO(8)$, which preserve an extra $U(1)$ that requires 
the SM fermions to couple to only one Higgs doublet. This avoids 
FCNC problems from the Higgs sector since the other PNGB doublet 
becomes inert. The  $Z\bar{b}_Lb_L$ issue is also 
coped with when all the fermions are embedded in the spinorial 
representations.  We also survey the different C2HDM that have been 
studied in the literature in our framework and compare and contrast
them with the new models presented here.

The paper is organized as follows. In Sections~\ref{SU5_coset}
-\ref{SU5_gauge_int} we state the problem and set the notation  
studying the coset and discussing the gauge interactions for 
$SU(5)/SU(4)\times U(1)$ and $SU(5)/SU(4)$. 
In Section~\ref{SU5_fermions_embedding} we discuss in detail the 
possible embeddings for the SM fermions in $SU(5)$ representations 
and their phenomenological consequences. 
In Section~\ref{SU5_potential}
we study the scalar potential in the $SU(5)$ case. 
In Section~\ref{metricnote} we introduce the gauge interactions and 
the tree level contribution to the $T$ parameter for different cosets.  
In Section~\ref{O9}  we analyze the $SO(9)/SO(8)$ gauge and 
fermion couplings to the Higgs sector. 
Finally we  conclude with some general observations.


\section{Non linear realization of the Higgs fields}\label{SU5_coset}

The general formulation of non linear representations (n.l.r.) is given 
in CCWZ. Here we introduce an alternative formulation which has also 
been used in models with a single composite  Higgs doublet like those 
quoted above.

We  begin with the study  the breaking patterns $SU(5)/SU(4)$ and $SU(5)/SU(4)\times U(1)$.
The coset space of the two cases is quite similar, although they
contain a different number of Goldstone modes. The Goldstone modes can
be parametrized as the matrix
\begin{eqnarray}
\Pi=\left(
\begin{array}{c|c}
 \frac{1}{\sqrt{20}}{\phi}_0 \mathbf{1}_4 & \begin{array}{c} \Phi
   \\ \tilde{\Phi} \end{array} \\ \hline
 \begin{array}{cc}
 \Phi^\dagger & \tilde{\Phi}^\dagger 
 \end{array} & -\frac{4}{\sqrt{20}}{\phi}_0
\end{array}\right),
\label{defPi}
\end{eqnarray}
where $\Phi$ and $\tilde\Phi$ are written in terms of two complex
$SU(2)_L$ doublets $\phi_1\, , \phi_2 $ as
\begin{equation}
\Phi\equiv\frac{\phi_1+i \phi_2}{\sqrt{2}} ~~\mbox{and}~~\tilde \Phi \equiv 
\frac{\tilde\phi_1+i \tilde\phi_2}{\sqrt{2}},
\label{phidef}
\end{equation}
with $\tilde \phi_i = \mathit{i} \sigma_2 \phi_i^*$ (see Appendix~\ref{app:fermrep}
for the conventions on the generators of the group).  Each $({\phi}_i
\, ,\tilde{\phi}_i )$ transforms as a ${\bf (2\,, 2)}$ of $SU(2)_L
\times SU(2)_R$, while the $9^{th}$ Goldstone boson, the singlet
${\phi}_0$, is present only in the $SU(5)/SU(4)$
coset. Correspondingly, the Goldstone boson matrix for the symmetric
coset $SU(5)/SU(4)\times U(1)$ is obtained for $ {\phi}_0 = 0$. 
The matrix $\Pi$ parametrizes the elements of the cosets in terms of the
PNGB fields.

To write the {n.l.r.} of the Higgs fields, it is convenient to choose 
a specific direction for the vacuum $\Sigma_0$. The breaking 
$SU(5) \rightarrow SU(4)$ can be parametrized either through the 
fundamental or the symmetric representation of $SU(5)$,
\[
{\bf 5}:~~
\Sigma_0^T = (0\  0\ 0\ 0\  1) 
\qquad {\bf 15}: ~~\Sigma_0 = \mathrm{diagonal}(0\  0\ 0\ 0\  1)
\qquad [SU(5)/SU(4)],
\]
while $SU(5) \rightarrow SU(4)\times U(1)$ can be parametrized through the
adjoint representation,

\[ {\bf 24}:~~
 \Sigma_0 = \frac{1}{\sqrt{20}}
\mathrm{diagonal}(1\  1\ 1\ 1\  -4) \qquad [SU(5)/SU(4)\times U(1)].
\]
The expressions for the n.l.r. in terms of the Goldstone bosons  in the 
two cases are given by
\begin{eqnarray}
 {\bf 5}: & \Sigma= e^{\frac{i \Pi}{f}} \Sigma_0 \nonumber\\
 {\bf 15}: & \Sigma= e^{\frac{i \Pi}{f}} \Sigma_0 e^{\frac{i \Pi^T}{f}}\nonumber\\
 {\bf 24}: & \Sigma= e^{\frac{i \Pi}{f}} \Sigma_0 e^{\frac{-i \Pi}{f}}
\end{eqnarray}
where the appropriate vacuum $\Sigma_0$ must be used and $f$ is the
scale at which the breaking occurs.

The n.l.r. is obtained in terms of a 5-components unit vector $u$ as 
follows. In the $SU(5)/SU(4)$ case,
$u$ reads
\begin{equation}\label{U5}
 u=e^{-i \frac{3 }{5}\frac{\phi_0}{f}}\;\left(
\begin{array}{c}
i \frac{\sin\left( \frac{ \varphi}{f}\right)}{\varphi} \begin{pmatrix}
  \Phi \cr \tilde\Phi\end{pmatrix}
  \\ \cos\left(\frac{\varphi}{f}\right)+i{\phi}_0
  \frac{\sin\left(\frac{\varphi}{f}\right)}{\varphi}
 \end{array}\right)= e^{\frac{i \Pi}{f}} \begin{pmatrix}
 0 \cr 0 \cr 0 \cr 0 \cr 1\end{pmatrix} = e^{\frac{i \Pi}{f}} u_0\,, 
 \qquad u^\dagger u=1\;,
\end{equation}
where $\varphi=\sqrt{|\Phi|^2 + |\tilde\Phi|^2 +{\phi}_0^2}$ and 
\begin{equation}
 \Sigma = u~~[{\bf 5}], ~~~\Sigma = u u^T~~[{\bf 15}]
\end{equation}

In the $SU(5)/SU(4)\times U(1)$ case, with ${\phi}_0 = 0$,
\begin{equation}\label{U24}
 u=\left(
\begin{array}{c}
i \frac{\sin\left( \frac{ \varphi}{f}\right)}{\varphi} \begin{pmatrix}
  \Phi \cr \tilde\Phi\end{pmatrix}
  \\ \cos\left(\frac{\varphi}{f}\right)
 \end{array}\right) = e^{\frac{i \Pi}{f}} u_0 , \qquad u^\dagger u=1\;,
\end{equation}
where now $\varphi=\sqrt{|\Phi|^2 + |\tilde\Phi|^2}$. In this case the
$\Sigma$ matrix reads
\begin{equation}
 \Sigma= -\frac{\sqrt{5}}{2}\left(u u^\dagger -\frac{{\bf 1}_5}{5}\right)\;.
\label{sigma24}
\end{equation}

Because $\varphi$ is invariant under the preserved $SU(4)$, $u$ linearly
transforms as a ${\bf 4 + 1}$, like the Goldstone fields in Eq.~\ref{defPi}. 
For the $SU(5)/SU(4)\times U(1)$ case, the $U(1)$ invariance
is obtained  by the multiplication of a phase defined by the transformation 
of the Goldstone bosons: $u \rightarrow gue^{i\theta}$, with 
$\theta=\theta(g, \phi^i )$. This corresponds to the transformation of 
$e^{\frac{i \Pi}{f}}$ defined in CCWZ  once the $SU(4)$ invariance of  
vacuum is taken into account. This phase disappears in the 
realization $\Sigma$ as defined in Eq.~\ref{sigma24}.

It is useful to formulate the Lagrangian for the PNGB fields and 
their interactions\footnote{For a review of the more usual 
CCWZ  approach in the context of EWSB see, e.g., 
\cite{Contino:2011np}. } in terms of $u$. First note that  
$\Sigma \Sigma^{\dag}$ are projectors  along the vector $u$ 
for the $\mathbf{5}$ and the $\mathbf{15}$ vacua. 
It implies that the insertions of the $\Sigma$ field in the different terms 
of the effective Lagrangian are quite limited. In practice one has only 
to introduce the minimal number of $u$'s needed to ensure the 
$SU(5)$ invariance. This is equivalent to the insertion of the unitary 
matrix $U=e^{\frac{i \Pi}{f}}$ in the CCWZ formalism. However, 
the breaking by a $\mathbf{15}$ preserves an additional parity 
symmetry: $u \rightarrow -u$. Therefore the u's must come by pairs, 
hence as $\Sigma = u u^T$.  For the $\mathbf{24}$ case, let us 
remark that the unit matrix in Eq.~\ref{sigma24} is irrelevant as it 
does not introduce any new operator in the effective Lagrangian. 
Therefore we henceforth replace Eq.~\ref{sigma24} 
by $\Sigma = uu^{\dag}$, which is a projector, $\Sigma^2
 =\Sigma$ . Therefore, the effective Lagrangian should only have 
 the minimal number of insertions of the $\Sigma$ field as needed 
 for the $SU(5)$  invariance in each interaction, much like in the previous 
 case. This  drastically simplifies the analysis of flavour changing 
 effects from Higgs couplings to fermions below. 

We now turn to the analysis of the Goldstone boson Lagrangians.
This can be done in terms of the $u$'s transforming in the 
fundamental representation of $SU(5)$, the simplest 
{n.l.r.}, from which the other ones can be built. Of course,
the Lagrangians for the Goldstone bosons depend only on the 
specific coset as shown in CCWZ. We easily determine 
the explicit Lagrangians in terms of the $u$'s by just imposing
the equivalence with the general CCWZ expression.
There are two $SU(5)$ invariants with two derivatives since 
$u^\dagger u=1$: 

\begin{eqnarray}
\partial_{\mu} u^{\dag}\, \partial^{\mu} u &=& u_0^{\dag} \partial_{\mu} 
U^{\dag} (UU^{\dag})\partial^{\mu} Uu_0  \nonumber \\
u^{\dag}\partial_{\mu}u  \,\partial^{\mu}u^{\dag} u&=& u_0^{\dag} U^{\dag} 
\partial_{\mu} U u_0\,u_0^{\dag}\partial^{\mu} U^{\dag} Uu_0
\label{invariant}
\end{eqnarray}
where $U=e^{\frac{i \Pi}{f}}$. Notice the projections along the $u_0$
direction. The element of the $SU(5)$ algebra $U^{\dag} \partial_{\mu}U$ 
can be expanded as, $id_{\mu}^i T^i  +  iE_{\mu}^a T^a
+ (d/E)_{\mu}^0 T^0$, where $T^a$ are the generators of $SU(4)$, 
$T^0$ the one of the $U(1)$ and $T^i$ those of the {coset}. The 
component along the $U(1)$ generator is $d_{\mu}^0$ for 
 $SU(5)/SU(4)$, and $E_{\mu}^0$  for $SU(4)\times U(1)$. The 
 $d_{\mu}$'s transform linearly under local $SU(4)$ and 
 $SU(4)\times U(1)$, respectively, while the $E_{\mu}$'s transform 
 as gauge fields \cite{Coleman:1969sm, Contino:2011np}.

Replacing these expression in Eq.~\ref{invariant} and projecting along 
$u_0$,  one gets for the two invariants two different expressions in terms
of $d_{\mu}^i d^{\mu i}$ and $E_{\mu}^0 E^{\mu 0}$. However, since 
$u_0$ is a $SU(4)$ singlet, the correspondent $E_{\mu}^a E^{\mu a}$ 
terms are absent in both the global invariants in Eq.~\ref{invariant}, while the 
terms $d_{\mu}^i$ transforming as $\mathbf{4+\bar{4}}$ are absent from 
the quartic one. Then, the CCWZ Lagrangian, ${\cal L}_{PNGB} = 
\Sigma_i d_{\mu}^i d^{\mu i}$ is given in each case by the combination
that eliminates or consistently normalizes the singlet term, resulting in 
local $SU(4)$ invariant two-derivative PNGB Lagrangians. They are as follows:
\begin{eqnarray} \label{metric}
 {\cal L}_{PNGB}  = &f^2 \left(\partial^{\mu} u^{\dag}\partial_{\mu} u
 - u^{\dag}\partial_{\mu} u\, \partial^{\mu} u^{\dag}u \right) & 
 [SU(5)/SU(4)\times U(1)]\nonumber \\ 
 = &f^2 \left( \partial^{\mu} u^{\dag}\partial_{\mu} u
- \frac{3}{8} u^{\dag}\partial_{\mu} u\, \partial^{\mu} u^{\dag}u\right) & 
[SU(5)/SU(4)]  
\end{eqnarray}
Notice that in the first lagrangian in (\ref{metric}) there is only one invariant
under local $SU(4)$, while the second Lagrangian differs by the presence of 
a second one, $d^0d^0 = (5/8)u^{\dag}\partial_{\mu} u\, \partial^{\mu} u^{\dag}u$, since with our normalization of $T^0$, $T^0 u_0= - \sqrt{8/5}.$ 

It is worth remarking that the use of the $u$ variables and the projection
into the $u_0$ automatically eliminates most of the $E_{\mu}$'s, 
towards the construction of the locally invariant Lagrangians. This
procedure is generalized in section \ref{metricnote} to other cosets.
If useful, the corresponding expressions for the $\Sigma$'s 
can be obtained by using their expressions in terms of the $u$'s. 
In the next section we show that the anomalous contribution to the 
T-term is generally encoded in the quartic term in the $u$'s of the 
Lagrangian.


\section{Gauge interactions}\label{SU5_gauge_int}

\begin{table}[h!]
  \centering
 \begin{tabular}{c|c|c}
  Vertex & 2HDM & C2HDM \\
  \hline
  $hW^+ W^-/ZZ$     & $2 \frac{m_{W,Z}^2}{v}$ 	& $2 \frac{m_{W,Z}^2}{v} \sqrt{1-\e^2}$  \\
  \hline
  $h^2WW/ZZ$        & $\frac{m_{W,Z}^2}{v^2}$ 	& $\frac{m_{W,Z}^2}{v^2}\left(1-2 \e^2\right)$ \\  
  \hline
  $H^2WW/ZZ$        & $\frac{m_{W,Z}^2}{v^2}$ 	& $\frac{m_{W,Z}^2}{v^2} \re \sqrt{1-\e^2} $ \\
  \hline
  $A^2WW$           & $\frac{m^2_W}{v^2}$ 	& $\frac{m^2_W}{v^2} \re \sqrt{1-\e^2} $\\
  \hline
  $A^2 ZZ$     	    & $ \frac{m_Z^2}{2 v^2}$  	& $\frac{1}{2}\frac{m_Z^2}{v^2} \left(\re \sqrt{1-\e^2} - 2 \e^2 \re^2\right)$\\ 
  \hline
  $H^+ H^- W^+ W^-$ & $2 \frac{m_W^2}{v^2}$ 	& $ 2 \frac{m^2_W}{v^2} \left( \re \sqrt{1-\e^2} - 2 \e^2 \re^2 \right)$\\ 
  \hline
  \multirow{2}{*}{$H^+ H^- ZZ$} & \multirow{2}{*}{$2\frac{m_Z^2}{v^2} \left( 1- 2 \frac{m_W^2}{m_Z^2}\right)^2$} & $2\left[\frac{m_Z^2}{v^2} \re \sqrt{1-\e^2} -\right.$  \\
  & 				& $ \left.- 4 \frac{m_W^2}{v^2} \left(1-\frac{m_W^2}{m_Z^2}\right)\re^2\right]$ \\
  \hline
  $H^- W^+ A_\mu(H+i A)$    & $ -\frac{ge}{2}$ &  $-\frac{ge}{2} r_\epsilon^2$ \\ 
  \hline
  $H^- W^+ Z H$        & $\frac{g' e}{2}$  &  $\frac{g'e}{2} r_\epsilon^2$ \\
  \hline
  $H^- W^+ Z A$     & $i \frac{g'e}{2}$ & 
  $i \frac{g' e}{2}\re^2 \left[1-2 \kappa \e^2 \left(1+\frac{g^2}{g'^2}\right) \right]$  \\ 
  \hline
  $H^+ H^+ W^- W^-$     & $\times$ & $\kappa \frac{4 m_W^2}{v^2} \e^2 \re^2$ \\ 
  \hline
 \end{tabular}
\caption{Gauge interaction for the Higgs fields in the C2HDM case following from Eq.~\refe{gauge_int}, 
as opposed to the usual renormalizable 2HDM case. The Higgs fields definition is given in Eq.~\refe{phys_higgs}. 
We used the shorthanded notation $\epsilon=v/f$ and $r_\epsilon=\epsilon/{\rm arcsin}(\epsilon)$.}
\label{table24}
\end{table}

The gauge Lagrangian for the Goldstone bosons in the effective theory
can be written from Eq.~\ref{metric} introducing covariant
derivatives for the $u$ variables, 
$\partial u_i \rightarrow \partial u_i - i (\hat W u)_i$, with 
$\hat{W}_{\mu} = gW_{\mu}^+T_L^- + gW_{\mu}^+T_L^- + 
eA_{\mu}(T_L^3 + T_R^3) + g \sec{\theta}_W Z_{\mu}(T_L^3 -
\tan^2{\theta}_W T_R^3)$.  Therefore the coupling of the SM gauge 
bosons to the PNGB fields contains two algebraically different terms,
\begin{equation}\label{gauge_int}
u^{\dag}\hat{W}_\mu \hat{W}_\nu u - 
   \kappa \,u^{\dag}\hat{W}_\mu u\,u^{\dag} \hat{W}_\nu u\ ,
\end{equation}
and only the first one, quadratic in $u$,  is analogous to the SM. 

The gauge interactions following from the previous expression are given in Table~\refe{table24}, with 
Higgs fields defined as
\begin{equation}\label{phys_higgs}
 \phi_1 =\left(\begin{array}{c} G^+ \\ h+iG^0
                \end{array} \right) 
  ~~ 
  \phi_2 =\left(\begin{array}{c}
                 H^+ \\
		  H+iA
                \end{array} \right)\; .
\end{equation}
and $G^{0,+}$ the electroweak Goldstone bosons. It is
interesting to note that different couplings between the Higgs boson
and the W and Z vectors could be important to explain the cross-section
and branching ratios of the observed excess around $125\;{\rm GeV}$
\cite{Farina:2012ea}.

In momentum space, the gauge Lagrangian reads
\begin{equation}
 {\cal L}_{gauge} = \frac{1}{2} p^T_{\mu\nu} \left[ \Pi_0(q^2) {\rm
     tr}( \hat{W}_{\mu} \hat{W}_{\nu}) + \Pi_1(q^2) 
   (u^{\dag}\hat{W}_\mu \hat{W}_\nu u - 
   \kappa \,u^{\dag}\hat{W}_\mu u\,u^{\dag} \hat{W}_\nu u)\right]
\label{gauge}
\end{equation}
where $\kappa = 1$ or $3/8$ as in Eq.~\ref{metric} and $\Pi_0(q^2) $ and 
$ \Pi_1(q^2)$ are form factors. Of course, this Lagrangian is not 
invariant under $SU(2)_L\times SU(2)_R  \subset SU(4)$  because only 
the SM subgroup is gauged. In 2HDM's,  only the alignment between 
the two doublets in the vacuum ensures a custodial $SU(2)$  to protect 
the $T$ parameter. 

Indeed, in both the cases under consideration the quartic term 
spoils the custodial symmetry already at tree level. To investigate this 
crucial point, let us keep in Eq.~\ref{gauge} only the components of $u$ 
with electric charge $T_L^3 + T_R^3 = 0$, namely, $u_2$ and $u_3$ 
with the conventions  defined in Appendix~\ref{app:fermrep}. The embedding 
of the $SU(2)_L \times U(1)_Y$ generators in $SU(2)_L\times 
SU(2)_R \subset SU(4)$ is uniquely defined by \rep{4} = \rep{(2,2)}, 
so that the masses of the gauge bosons are obtained from
Eq.~\ref{gauge} as follows:
\begin{equation}
{\cal L}_0 = f^2 \left( \langle |u_2|^2 \rangle + \langle |u_3|^2
 \rangle \right) \frac{g^2}{2}\left( W^+_\mu W^-_\mu + \frac{1}{2
   \cos^2\theta_W} \left(1-\kappa\,\frac{( \langle |u_2|^2 \rangle - \langle
   |u_3|^2\rangle)^2}{ \langle |u_2|^2 \rangle + \langle |u_3|^2
   \rangle}\right) Z_\mu Z_\mu\right).
\label{rho5}
\end{equation}
\noindent
The additional contribution to the Z mass  only vanishes for 
$\langle |u_2|\rangle= \langle|u_3|\rangle$, which is precisely the 
condition for the alignment of the two doublets in the vacuum and 
the conservation of the custodial symmetry. Since the PNGB potential 
is generated at loop level, where $SU(2)_L\times SU(2)_R \subset 
SU(4)$ is explicitly broken,  it cannot generically guarantee such a 
degeneracy for the solution, so that a large contribution to the $T$ 
parameter should be present already at tree level. This would impose 
an unnatural hierarchy between the two relevant scales, $f \gg v$ 
(slightly less for $\kappa = 3/8$).

In principle: {\it i}) the custodial symmetry could be effective in a 
region of the parameter space for the solution of the 
equations of motion,  and
{\it ii}) the contributions to the $T$ parameter (both at tree and loop 
level) from the scalar sector can be partially compensated by
fermionic contributions \cite{Barbieri:2007bh}. 
Nevertheless, the mechanism for this 
approximate cancellation must be natural.

The conditions to avoid Higgs misalignments in composite 2HDM have
been exhaustively discussed in \cite{Mrazek:2011iu}, in particular, by the
use of additional discrete symmetries. They were able to build a model
with one Higgs coupled to the SM fermions and an inert Higgs that 
does not contribute to the gauge boson masses. The natural 
discrete symmetry for $SU(5)$ is charge conjugation ($\phi_1 \rightarrow \phi_1$ and $\phi_2 \rightarrow -\phi_2$,
see Eq.~\ref{phidef}).
To investigate this
point, as much as the mostly important flavour issues, we now turn to
a study of the SM fermion embedding and their couplings to the 
two Higgs bosons.


\section{Fermion Embeddings}\label{SU5_fermions_embedding}

As usual in Composite Higgs models, the coset structure does not fix 
the embedding of the fermions in $G$ representations. Moreover, 
different choices will in principle generate different terms in the 
Coleman-Weinberg potential. However, some general considerations 
allow to make the model theoretically consistent and phenomenologically 
viable. Some of the important issues are:

\begin{enumerate}
 \item LEP has measured the $Zb_L\bar{b}_L$ coupling with high
   precision, finding results in agreement with the SM prediction at
   nearly $0.25\%$ level\footnote{ Recent 2 loop calculations indicates that
   the agreement between SM and experimental values of the $Zb_L\bar{b}_L$ coupling  
   may not be as good as previously expected \cite{Freitas:2012sy}. }. As shown in \cite{Agashe:2006at}, a custodial
   symmetry can protect this coupling from large contributions due to
   composite states. The argument boils down to the conclusion that
   the left handed quark doublet (at least the third generation)
   should be embedded in a ${\bf (2,2)_{2/3}}$ representation of $SU(2)_L
   \times SU(2)_R \times U(1)_X$ in order to protect this coupling from 
   getting too large corrections.
\item If the SM fermions can be embedded into $G$ representations in 
a unique way, then the problem of FCNC is considerably reduced 
\cite{Gripaios:2009pe,  Agashe:2009di}. There are cases where 
this can be achieved by the addition of discrete symmetries 
\cite{Mrazek:2011iu}. In the $SO(9)/SO(8)$ model described 
below this is done by a $U(1) \in SO(8)$.
\item In 5d models of gauge-Higgs unification that are perturbative dual to the 
strongly  coupled 4d model, where the PNGB are the  $5$-th component 
of gauge bosons, the fermions of both chiralities should be in the same 
representation of the gauge group (global group, from the 4d point of 
view)\footnote{In principle, this requirement can be circumvented by 
introducing extra fermion multiplets and mass terms in the IR brane . 
However, this is a potential source of FCNC, so that we discard this 
assemblage in our discussion.}. However, this is not a stringent 
consideration for the purely 4d model, and we will actually explore 
some of the advantages of using LH and RH fermions in different 
representations.
\end{enumerate}

In terms of the SM fermions, the Lagrangian {in momentum space reads
\cite{Agashe:2004rs}}
\begin{eqnarray}
 {\cal L}_{fermion} &=& \overline{q}_L \; \slashed{p} \left[ \Pi_0^q +
   \Pi_1^q g_{q_L}(u_i) \right] q_L + \nonumber\\ && + \overline{t}_R
 \; \slashed{p} \left[ \Pi_0^u + \Pi_1^u g_{t_R}(u_i) \right] t_R +
 \overline{b}_R \; \slashed{p} \left[ \Pi_0^d + \Pi_1^d g_{b_R}(u_i)
   \right] b_R + \nonumber\\ && + f\overline{t}_R \left[M_1^u
   g_{tq}(u_i) \right] q_L +f\overline{b}_R \left[M_1^d g_{bq}(u_i)
   \right] q_L
\label{fermion_epan} 
\end{eqnarray}
where the $\Pi_i(Q^2)$ and $M_i{Q^2}$ are form factors controlled by the strong
sector and the $g(u_i)$ functions of the Higgs fields. These polynomials
are defined by the possible insertions of the n.l.r. $u$'s, or $\Sigma$'s
in the fermion bilinear terms $\overline{\Psi}_A^i g^{ij}(u) \Psi_B^j$,
with $A,B=L,R$ that are $SU(5)$ invariants. 
The Lagrangian is obtained by restricting the fermion multiplets to their SM components.

For simplicity, in what follows we will limit ourselves to the lower dimensional
representations of $SU(5)$, {\it i.e.} ${\bf 5}$ and ${\bf 10}$,
although  higher dimensional representations in
principle can be used. In terms of $SU(2)_L \times SU(2)_R$ they read
\begin{eqnarray}\label{irrepsdecomposition}
 \rep{5} &=& \rep{(2,2)}+\rep{(1,1)} \nonumber\\
 \rep{10} &=& \rep{(2,2)}+\rep{(3,1)} + \rep{(1,3)}
\end{eqnarray}

Let us first discuss the case in which both the left (LH) and right (RH) 
fermions are embedded in the fundamental representation. To 
accommodate hypercharge, each RH singlet must be embedded in 
an independent $\rep{5}$, with appropriate X-charge. This implies that, 
by $U(1)_X$ invariance, $q_L$ must be associated to two $\rep{5}$, 
respectively $\Psi_L^t$ and $\Psi_L^b$. In a 5D picture, this is achieved 
decoupling the additional degrees of freedom through a mass term in 
the brane \cite{Contino:2006qr, Gripaios:2009pe}. However, since 
$b_L$ does not belong to $\rep{(2,2)_{2/3}}$, the $Zb_Lb_L$ vertex 
is not protected. In models with only one Higgs doublet, this is not 
necessarily a shortcoming since the breaking is proportional to the 
relatively small bottom Yukawa coupling \cite{Gripaios:2009pe}.  In 
2HDM this is a cause for concern for large 
$\tan\beta$.

Turning to the case in which both the LH and RH fermions are 
embedded in a $\rep{10}$, the RH singlets must be in a unique 
${\bf (1,3)}\in \rep{10}_{2/3}$, with $T_{3R}=0$ for $u_R$ and 
$T_{3R}=-1$ for $d_R $.  As a consequence, also $q_L$ is
embedded in a unique representation and $Zb_Lb_L$ is protected.
This is the $SU(5)$ counterpart of the choice of the real $\rep{10}$ 
for the fermions in the $SO(5)/SO(4)$ model.

In both cases, there is only one possible embedding of the SM fermions 
in the $SU(2)_L \times SU(2)_R$ representations. Another 
important feature: for all $\overline{\Psi}_A^i g^{ij}(u) \Psi_B^j$,
we find  $g^{ij}(u) \propto uu^{\dag} = \Sigma$. Indeed, 
$\Sigma$ is a projector, $\Sigma^2=\Sigma$, and 
$\Sigma \Psi_{L,R}^i \Sigma =0$ for the SM in a $\rep{10}$. 
The expressions for the $g$ functions, as defined in Eq.~\refe{fermion_epan}, 
are collected in Table~\ref{gfunc} for both $SU(5)/SU(4)$ and 
$SU(5)/SU(4)\times U(1)$, choosing the appropriate $u$ in 
 Eqs.~\ref{U5}-\ref{U24}. A crucial attribute of the mass terms is that the up 
 quarks get masses only from one combination of the two Higgs, while
 the down quarks couple to the orthogonal one. Therefore there is no 
 flavour violation from the Higgs couplings.

 \begin{table}[tb]
\begin{center}
  \begin{tabular}{c|cc}
    & $\rep{5}$ & $\rep{10}$ \\ 
\hline
  $g_{q_L}$ & $|u_3|^2+ |u_4|^2,  |u_1|^2+ |u_2|^2$ &  $\begin{pmatrix}
 |u_5|^2 + |u_3|^2 &  u_4^* u_3 \cr
 u_4 u_3^* & |u_5|^2 + |u_4|^2
\end{pmatrix}$ \\
  $g_{t_R}$ & $|u_5|^2$ & $|u_1|^2+|u_2|^2+|u_3|^2+|u_4|^2$ \\
  $g_{b_R}$ & $|u_5|^2$ & $|u_3|^2+|u_4|^2$\\
  $g_{tq}$ & $u_5 \begin{pmatrix} u_3 \cr u_4,\end{pmatrix}^\dagger$ & $u_5^* \begin{pmatrix} u_1 \cr
   u_2 \end{pmatrix}^T (i\sigma_2)$ \\
  $g_{bq}$ & $u_5 \begin{pmatrix} u_1 \cr u_2,\end{pmatrix}^\dagger$ & $u_5^* \begin{pmatrix} u_3 \cr u_4 \end{pmatrix}^T (i\sigma_2)$\\
  \end{tabular}
\end{center}
\caption{\label{gfunc} Couplings between the SM fermions and the $u$ 
variables, as defined in Eq.~\refe{fermion_epan}, according to the $SU(5)$ 
representation in which they are embedded. The result can be applied 
both to $SU(5)/SU(4)$ and $SU(5)/SU(4)\times U(1)$ choosing the 
appropriate expression for $u$. In the case of the $\rep{5}$, 
the first value for $g_{q_L}$ refers to $q_L^t$, the second to $q_L^b$.}
 \end{table}

In all the previous cases, a 5-dimensional gauge-Higgs unification
can be  consistently constructed.  However, from a purely 4d perspective,
the embedding 
of LH and RH fermions in different representations make sense, so that 
we will now briefly comment also on this possibility. Since we are confining 
ourselves only to the $\rep{ 5}$ and $\rep{10}$, only for $SU(5)/SU(4)$ the 
embedding in different representations is possible. The simplest 
possibilities are given by:
\begin{description}
 \item[(a)] $(\Psi_L,\Psi_R)\in \rep{(5, 10)}$,  with invariant 
 ${\bf 5}_\Sigma^T \overline{\bf 10}_R {\bf 5}_L$. Note that left doublet 
 can be embedded in the $\rep{(2,2)_{2/3}}$, protecting $Zb_L\bar{b}_L$.
 
 \item[(b)] $(\Psi_L,\Psi_R)\in \rep{(10, \overline{10})}$, with
   invariant $\epsilon^{ijkmn} ({\bf 10}_R)_{ij} ({\bf 10}_L)_{km}
   ({\bf 5}_\Sigma)_n$.  Since $b_L = ({\bf 10}_L)_{54}$ and $b_R
   =(\overline{\bf 10}_R)_{34/43}$, a mass term $\bar{b}_R b_L$ is
   forbidden and the bottom mass can only be generated at higher
   order, explaining in a natural way the bottom/top mass gap.
   However in both this cases there is no direct couplings between the
   fermions and the $u_5$.

 \item[(c)] $(\Psi_L,\Psi_R)\in \mathbf{(5, \overline{5})}$, which
leads to a phenomenology similar to the ${\bf (5,5)}$ case up to a
phase in the Yukawa interactions. 
 \end{description}

Let us comment on how the flavor problem is solved in our case.  
As pointed out in~\cite{Mrazek:2011iu}, the problem is twofold: 
on the one hand, one has to face the usual 2HDM problem of 
ensuring Minimal Flavor Violation (MFV) in the Yukawa sector;  
on the other hand, there could be more  invariant Yukawa couplings
of the fermions to the Higgs n.l.r. and a priori there is no reason to have 
alignment between the different flavor matrices defining different couplings 
~\cite{Agashe:2009di}. In the $SU(5)$ models under consideration, 
the flavour mixing is solely given by the mass matrices so that MFV is realized 
because 

\noindent
(\textit{i}) the representations considered, \rep{5} and \rep{10}, 
are small enough to allow only one embedding of the up and down quarks, 
as displayed in Eq.~\ref{irrepsdecomposition};

\noindent
(\textit{ii}) all the fermions only couple to one $\Sigma$, so that their couplings
to the Higgs are flavour diagonal.

Hence, there is no tree-level flavour changing from the Higgs sector. 
Unfortunately, the $t$-quark, instrumental in EWSB, couples to $u_3$,
which is a combination of the real part of a Higgs field and the imaginary 
part of the other, pushing for the misalignment for their vev's. To 
point out this serious problem, we briefly discuss the PNGB potential
in the next section.


\section{Scalar Potential}\label{SU5_potential}

The Higgs fields arise as Goldstone modes, so  the 
shift symmetry  forbids a tree level potential. However, the global 
symmetry is explicitly broken once the SM group is gauged and the fermions 
are coupled to the Higgs system, so that the Goldstone modes get a 
Coleman-Weinberg potential. The potential can be computed with standard techniques, and may be written in general as
\begin{eqnarray}
 V &=& V_\mathrm{gauge} + V_\mathrm{fermion}, 
\label{pot}
 \end{eqnarray}
In terms of the $u$ variables, and to lowest order in $g^2$ and $g'^2$, the gauge part of the potential for the neutral components reads
\begin{eqnarray}
V_\mathrm{gauge} &=& \frac{3}{2}\int \frac{d^2Q}{16\pi^2}
\frac{\Pi_1}{\Pi_0}\left( \frac{3g^2+g'^2}{4} \left(|u_2|^2 +|u_3|^2 \right) + D^AD^A  \right)
\label{potgauge}
\end{eqnarray} 
with $D^A= g_A u^\dag T^A u$. This term is the additional contribution 
arising from the terms quartic in $u$ in the gauge part of the Lagrangian, 
as given in Eq.~\refe{metric}, and closely resembles the supersymmetric 
case (without $B\mu$ term).

The fermionic contribution $V_\mathrm{fermion}$ in Eq.~\refe{pot} can be 
simplified in the reasonable limit in which $b_R$ has a relatively small 
coupling to the strong sector .With this assumption, and to lowest order 
in the ratios of structure functions $\Pi^a_1/\Pi^a_0$, defined in 
Eq.~\refe{fermion_epan}) the potential for the neutral components results:
\begin{eqnarray}
V_{\rm fermion}^{(\rep{10})} 
&=& \int \frac{d^4Q}{16\pi^2}  \left[\left( \rho^{q_3} + \frac{1}{2}  \rho^{t_R} \right) 
\left(|u_2|^2+ |u_3|^2 \right) + \left( \rho^{q_3} - \frac{f^2}{Q^2}Y_t^2 \right) |u_3|^2 \right.  \nonumber\\
& &\left. + \frac{1}{2} \rho^{t_R} \left(|u_2|^2 + |u_3|^2 \right)^2 + 
\frac{f^2}{Q^2}Y_t^2  \left( |u_2|^2+ |u_3|^2 \right)|u_3|^2  + . . . \right]
\label{potferm10}
\end{eqnarray}
for fermions embedded in a $\rep{10}$, and
\begin{eqnarray}
V_{\rm fermion}^{(\rep{5})}  
&=& \int \frac{d^4Q}{16\pi^2} \Bigg[ \rho^{q_3} \left(1-|u_2|^2 - |u_3|^2 \right) + \rho^{t_R}  
\left(|u_2|^2 + |u_3|^2 \right) 
\nonumber\\
& & \left. - \frac{f^2}{Q^2}Y_t^2 |u_3|^2  \left(1-|u_2|^2 - |u_3|^2 \right) 
+  \rho^{t_R} |u_2|^2 |u_3|^2 + . . . \right]
\label{potferm5}
\end{eqnarray}
for fermions embedded in a $\rep{5}$, where the 
integrands are the ratios,
\begin{equation}
\rho^f (Q^2) = \frac{\Pi_1^f}{\Pi_0^f}\, , \  (f=t_R,\,b_R,\, q_3) \quad \mathrm{and}
\quad Y_t(Q^2) = \frac{|M_1^t |^2}{\Pi_0^{q_3}\Pi_0^{t_R}}
\end{equation}

Of course, a quantitative discussion of the Higgs vev and the spectrum is 
pointless in presence of so many  unknown form factors. However, 
some comments are now in order:

\begin{enumerate}
\item To return to the Higgs alignment issue, notice that the 
important negative contribution in the potential corresponds 
to the $t$-quark Yukawa coupling to $u_3$.  Hence it  drives 
the vacuum through the $u_3$ direction which breaks the 
custodial symmetry as already stressed. The $b$-quark 
coupling to $u_2$, neglected in (\ref{potferm10}, 
\ref{potferm5}), is too small to compete. As already quoted 
above, Higgs alignment can be enforced in composite 2HDM 
by discrete symmetries~\cite{Mrazek:2011iu} that allow to 
restrict the couplings of SM fermions to a single Higgs 
in the case of the $SU(4)/SU(2)\times SU(2)\times U(1)$
model. In our case, the interesting property of the models is 
precisely the automatic uniqueness of these couplings and 
also the fact that quarks of different charges interact with 
different Higgs fields. Therefore, we cannot impose a 
symmetry that, at most, would kill our single Yukawa coupling.
We have to rely on a partial compensation mechanism.

\item Since the custodial symmetry is essentially broken by the
  Higgs coupling to $q_L$ (see Appendix \ref{app:fermrep}) that 
  explicitly breaks $SU(2)_R$, one could try to improve the 
  situation introducing custodial fermions in the \rep{(2,2)} of 
  $SU(2)_L \times SU(2)_R$~\cite{Pomarol:2008bh} to alleviate 
  the constraint from the $T$-parameter. Interestingly, they might 
  also contribute through loops to increase $BR(h\rightarrow 
  \gamma \gamma)$~\cite{Azatov:2011qy}, in agreement with 
  the recent observations at the LHC. They naturally provide an 
  interesting phenomenological set up, detailed study of which 
  is beyond the mandate of this paper.

\item For $SU(5)/SU(4)$, the $\varphi_0$ component of $u$ is 
  a main concern too, since it is  the PNGB associated to a 
  abelian symmetry and, as such, an axion-like scalar subject to 
  strict experimental bounds.It always appears in the potential as 
  $|u_5|^2 =1-\sin^2\left(\frac{\varphi}{f}\right)\frac{|\Phi|^2+
  |\tilde{\Phi}|^2}{\varphi^2}$  The fact that $\varphi_0$
  only appears as a function of $\varphi$ implies that there is always
  a massless particle in the spectrum. Indeed, if we consider the
  simplified case in which $u$  has only one non vanishing 
  component, since $V=V(|u|^2)$, at the minimum the mass 
  matrix in the $(u,\varphi_0)$ basis has the form
      \[ M^2=
       \begin{pmatrix}
        1 &  \frac{\partial |u|^2}{\partial \eta} \cr
        \frac{\partial |u|^2}{\partial \eta} &  \left(
        \frac{\partial |u|^2}{\partial \eta} \right)^2
       \end{pmatrix}\partial^2 V / \partial |u|^2
      \]
   and has a zero eigenvalue. The result trivially generalizes 
   to more $u$ components. This problem is independent of 
   the choice of fermion embeddings. It may be solved introducing 
   two or more right handed neutrinos. Consider $N_R^{1,2}$, two 
   neutrino that couple to two different strong sector operators, having
   representations $N_R^{1,2} = 5 + \zeta^{1,2} \bar{5}$. In this case
   a potential for the singlet field is generated at the leading order
   by the coupling $N_R^{1}\bar{N}_R^{2} \Sigma+ h.c.$. This might
   lead to interesting phenomenological consequences~\footnote{In 
   particular, the breaking of lepton number by the strong sector should 
   be treated carefully \cite{Frigerio:2011zg}. We thank J. Serra for 
   calling our attention on this point.} that are beyond the scope of the 
   present paper.

\end{enumerate}

\section{Other composite 2HDM's: non-linear realizations, PNGB Lagrangian and $T$ parameter} \label{metricnote}

As pointed out in the previous section, the presence of a term quartic in $u$ in 
the metric generates a contribution to the $T$ parameter already at tree level. 
This despite the presence of a large residual global symmetry $H \subset G$ containing the custodial symmetry as a subgroup.

In this section we display the PNGB Lagrangian for the different cosets leading to 
2HDM's.  We consider cosets $G/H_1 \times H_2$ with $H_2 = \emptyset, \, 
U(1),\, SU(2)$ and we generalize the approach in Section~\ref{SU5_coset}. As already noticed 
above for $G=SU(5)$,
it is useful to introduce as coset coordinates $p$ orthogonal unit vectors 
$u^{\alpha} $ ($\alpha = 1, \dots, p$) in the fundamental representation 
$\mathbf{N}$ of $G$ 
whose scalar products satisfy: 
\begin{equation}
u_{\alpha}^{\dag} u^{\beta} = \delta_{\alpha}^{\beta}
\quad \quad \mathrm{tr}\, u^{\dag} u \equiv 
\sum_{\alpha} u_{\alpha}^{\dag} u^{{\alpha}}= p
\end{equation}
where $p=1$ for $H_2 = \emptyset, \, U(1)$ and  $p=2$ for 
$H_2 = O(2),\, SU(2)$. They are functions of the PNGB fields, $\xi^A$ defined by
\begin{equation}\label{u_def}
u (\xi) = e^{i\Pi (\xi)} u(0)   \quad \quad   \Pi (\xi)= \sum_A \xi_A T^A
\end{equation}
where $ \xi_A $ are the PNGB fields and $T^A $ are the coset charges, 
while $u(0)$ corresponds to the vev('s) of one or more n.l.r. of the 
Higgs in fundamental representation(s) needed for the breaking. $G$ 
acts as $u \rightarrow guh_2^{\dag}(g,\xi )$. In particular, as already 
displayed above for $G=SU(5)$, n.l.r. transforming as larger 
irreducible transformations are obtained with  $\Sigma = uu^{\dag}$ 
for the adjoint representation, $\Sigma = uu^{T}$ for the symmetric, etc. 
Notice the contraction of the $p$ indices associated to the action
of $H_2$.

For $G=SU(5)$ the results are already given in (\ref{metric}). For the other 
cosets associated to C2HDM's, we just display the PNGB Lagrangians that 
are obtained along the  same lines.  For simplicity, the Lagrangians 
are arbitrarily normalized.\footnote{Actually the normalization can be fixed
the end, having  fitted to data the expressions for the gauge boson masses 
and normalized the Higgs boson kinetic term.} There are two invariants 
with two derivatives, one quadratic, and the other quartic in the $u$'s. 
By replacing $U^{\dag} \partial_{\mu}U = id_{\mu}^i T^i  +  iE_{\mu}^a T^a$ 
as explained in section \ref{SU5_coset}, we recall that by $H_1$ 
invariance of $u(0)$, the quadratic term becomes $\sum d_{\mu}^{i\,2}
+c \sum' E_{\mu}^{a\,2}$, with $\sum'$ restricted to the $H_2$ charges, while
only the last term is present in the quartic term (with another coefficient). 
By the appropriate combination of the two invariants, we thus obtain the 
CCWZ Lagrangian,  $\sum d_{\mu}^{i\,2}$,  in terms of the $u$'s.

First, consider $SO(6)/SO(4)\times SO(2) \equiv 
SU(4)/SU(2)\times SU(2)\times U(1)$ and let us first consider the construction
in terms of two {\bf 4} representations of $SU(4)$ satisfying the previous
rules, namely, $u^{\alpha}_i \ (\alpha=1,2\,; i=1,\dots ,4)$, $H_2=SU(2)$. 
Then,
\begin{equation}{\label{4221}}
 {\cal L}_{PNGB}  = f^2  \mathrm{tr}\,\left(\partial^{\mu} u^{\dag}\partial_{\mu} u
 - u^{\dag}\partial_{\mu} u\, \partial^{\mu} u^{\dag}u \right) \quad \quad 
 (SU(4)/SU(2)^2\times U(1))  
\end{equation}

But, this can be also written in terms of two (real) vectors of $SO(6)$, denoted 
$v^{\alpha}_a \ (\alpha=1,2\,; a=1,\dots ,6)$, $H_2=SO(2)$ to obtain, 
\begin{equation}{\label{642}}
 {\cal L}_{PNGB}  = f^2  \mathrm{tr}\,\left(\partial^{\mu} v^T\partial_{\mu} v
 - v^T\partial_{\mu} v\, \partial^{\mu} v^Tv \right) \quad \quad 
 (SO(6)/SO(4)\times SO(2)) 
\end{equation}
which, up to a normalization,  is the same as (\ref{4221}) if one replaces 
$v_a^{\alpha} = \bar{u}^{\alpha}\Gamma_a {u}^{\alpha}$, where $\Gamma_a$
are $SO(6)$ Dirac matrices.

The $Sp(6)/Sp(4)\times SU(2)$ model has some interesting 
properties.  With $H_1=Sp(4)$ and $H_2=SU(2)$, the n.l.r. can be 
displayed in terms of two $6$-vectors such that $u_2 = \Omega u_1^*$ 
leading to a triplet of quaternions as discussed in Appendix~\ref{app:Sp6}.
With the procedure described above we get
\begin{equation}\label{eq:Sp6_metric}
{\cal L} =f^2 \mathrm{tr}\left( \partial^\mu u^\dag \partial_\mu u - \partial^\mu u^\dag u 
u^\dag \partial_\mu u \right) \qquad \qquad (Sp(6)/Sp(4) \times SU(2))
\end{equation}

The  $SO(9)/SO(8)$ manifold is the 8-sphere, resulting in the corresponding
metric for the Lagrangian for the real 9-vector $u$,  \begin{equation}{\label{98}}
 {\cal L}_{PNGB}  = f^2  \mathrm{tr}\,\left(\partial^{\mu} u^T\partial_{\mu} u
\right) \qquad \qquad  (SO(9)/ SO(8))  
\end{equation}
Here, $H_1= SO(8) $ and, since there is no $H_2$, the quadratic terms 
already contains only  the PNGB part.  

The next step is to define the  embedding of the SM charges inside 
$H_1\times H_2$. Then, substituting the standard derivatives 
with the SM covariant derivatives, 
$\partial u_i \rightarrow \partial u_i - i (\hat W u)_i$, with 
$\hat{W}_{\mu} = gW_{\mu}^+T_L^- + gW_{\mu}^+T_L^- +
eA_{\mu}(T_L^3 + T_R^3) + g \sec{\theta}_W Z_{\mu}(T_L^3 -
\tan^2{\theta}_W T_R^3)$, the gauge boson masses and possible 
contributions to the $T$ parameter can be read from the previous 
expressions. All 2HDM's have $SU(2)_L\times SU(2)_R $ as a subgroup 
of the linear symmetry of the coset with the Higgs fields transforming 
in two $(\mathbf{2,2})$, so to define the wanted custodial symmetry. 
Let us take first $SU(2)_L\times SU(2)_R  \subset H_1$.  
The mass matrix for the gauge bosons can be written,
\begin{equation} {\label{M2AB}}
\frac{{\cal M}^2_{AB} }{g_A g_B} =
u^{\dag}\{T^A,T^B\} u - \kappa \,u^{\dag}T^Au\,u^{\dag}T^B u
\end{equation}
where $\kappa$ is given in the different ${\cal L}_{PNGB}$ above,
and where only the neutral components are retained, so that
$(T_L^3 + T_R^3)u = 0$. 
Since all the states are doublets or singlets of each $SU(2)$, 
$u^{\dag}\{T^A,T^B\} u = 1/2 \delta^{AB}
\sin^2 ( \varphi /f)$. The first term is analogous to the SM
gauge bosons mass term. The second term is easily calculated,
and gives the additional contribution to the $Z$-mass in (\ref{rho5})
and the corresponding anomalous contrition to the $T$-parameter.
Although the calculation is straightforward, we justify the results
from a group theoretical point of view in Appendix~\ref{app:D&T}. 

Up to the factor $\kappa$ the result is the same for all models where
the SM gauge symmetries are embedded in $H_1$ and that are 
strongly constrained by  $T$ unless the Higgs are aligned. The
exception are the models based on $SO(9)/SO(8)$ which have
only the quadratic part. In view of this success it deserves some 
limelight and it is further analysed in the next section.

The $Sp(6)/Sp(4)\times SU(2)$ models are special because, according 
to the embedding of the SM they can have enough symmetry to 
preserve the custodial symmetry that protects the$T$ 
parameter~\cite{Mrazek:2011iu}. Indeed, with the Higgs bosons in the 
\rep{(2,2)}, one of the factors in $SU(2)_L \times SU(2)_R$ must be
in $H_2=SU(2)$ and the other in $H_1=Sp(4)$. The latter can come 
from a maximal subgroup $SU(2)\times U(1)$ of $Sp(4)$ or from the
$SU(2) \times SU(2)$ one. Only the latter has sufficient symmetry
to protect $T$, since the two Higgs vev's become invariant under the
$SU(2)_C$ defined by the sums of the charges in the three $SU(2)$
factors. This is further discussed in Appendix~\ref{app:D&T} and the final
result is
\begin{equation}
M_W = M_Z \cos (\theta_W) = \frac{g f }{\sqrt{2}} \sin(h/f) \cos(h/f)
\end{equation}
where $h$ is the combined vev's of the two Higgs. Notice that 
higher order couplings of the Higgs sector to the gauge bosons are 
present at tree-level even if $\rho=1$ and, also, the factor  $\cos(h/f)$
with respect to the other models.

\section{A Composite Two Higgs doublet model based on $SO(9)/SO(8)$}\label{O9}

\noindent
We turn now to a composite 2HDM based on the symmetry breaking pattern $SO(9)/SO(8)$. 
As stressed in Section~\ref{metricnote} , there is no tree-level 
contribution to the $T$ parameter. The n.l.r.  is a 9- vector 
\begin{equation}\label{U9}
 u = e^{\frac{i \Pi}{f}} u_0 = \left(
\begin{array}{c}
\frac{\sin(\varphi/f)}{\varphi} \phi \\
\cos(\varphi/f)
\end{array}
\right),
\end{equation}
where $\varphi = \sqrt{\phi^T\phi}$, and $\phi$ the vector containing the 
eight NGBs {belonging to a $\rep{8_V}$ of $SO(8)$}.\footnote{We use 
here the conventions for the $SO(8)$ representations and subgroups from \cite{Slansky:1981yr}} 
The PNGB  Lagrangian has the single term in 
Eq.~\refe{98}. We need to embed the SM gauge group in $SU(2) \times SU(2)
\subset SO(8)$  to include the necessary custodial symmetry, and the two 
Higgses in the $\rep{8_V}$ transform as two $\rep{(2,2)}$. The selection
of these embeddings are given in Appendix~\ref{app:app_gen_O9}. 
There are two inequivalent choices that basically differ by  a $U(1) $ 
factor that differentiates between the Higgs doublets: 
\begin{itemize}
\item[{}] $\rep{8_V}=\rep{(2,2)}+\rep{(2,2)} \in SU(2)\times SU(2)\subset SO(8)$ 
\item[{}] $\rep{8_V}=\rep{(2,2)_{+1}}+\rep{(2,2)_{-1}} \in SU(2) \times SU(2) 
\times U(1) \subset SO(8)$
\end{itemize}
\noindent
In the first case, the fermions can couple to both Higgs doublets, which is a 
source of flavour changing. We do not investigated possible discrete
symmetries and concentrate on the second embedding where the
Higgs doublets have opposite $U(1) $ charges. 

To guarantee a correct electroweak charge assignment for the Higgs 
doublets, the additional $U(1)$ should not take part in the definition of the 
hypercharge, still given by $Y = T_L^3 + T_R^3$. With these assignments,
replacing the SM covariant derivative in Eq.~\refe{98}, one obtains the
masses of the gauge bosons 
\begin{equation}
 {\cal L}_m = \frac{1}{2} g^2 f^2 \sin^2(\varphi/f) \left( W^+_\mu W^-_\mu 
+ \frac{1}{2 \cos\theta_W} Z_\mu Z_\mu\right)\;,
\end{equation}
so that there are no tree-level restriction
on $v/f$, independently of the alignment between the two Higgs 
doublet in the vacuum.

Let us turn now to the fermions. Once the NGBs are assigned to the 
$\rep{8_V}$ of $SO(8)$ decomposing under $SU(2)\times SU(2)\times U(1)$ 
as $ \rep{(2,2)_{+1}}+ \rep{(2,2)_{-1}}$,  the other two eight-dimensional 
representations of $SO(8)$ transform under the  $SU(2) \times 
SU(2)\times U(1)$ subgroup as:  $\rep{8_s} = \rep{(1,1)_{+2}} +
\rep{(1,1)_{-2}} + \rep{(3,1)_{0}} + \rep{(1,3)_{0}}$ and $\rep{8_c} =
\rep{(2,2)_{-1}}+ \rep{(2,2)_{+1}}$. The assignments for $U(1)_X$ 
are parallel to those in Section~\ref{SU5_fermions_embedding}. Barring 
larger representations, there are two choices for the fermions, 
which are made explicit below
together with, in brackets,  their embeddings in $SO(9)\times U(1)_X$:

\begin{enumerate}
\item $t_L \in \rep{8_V} (\in \rep{9}_{2/3})$,  $b_L \in \rep{8_V} 
(\in \rep{9}_{-1/3})$, $b_R \in \rep{1}(\in \rep{1}_{-1/3})$,  
$t_R \in \rep{1}(\in \rep{1}_{2/3})$. A brane mass term allows for the
reconstruction of the $q_L$ doublet as sketched  in 
Section~\ref{SU5_fermions_embedding}. The Yukawa 
couplings  are the singlets in the products $\rep{8_V \times 
8_V \times 1} \in \rep{9 \times 9 \times 1}$. 
\item $q_L \in \rep{8_c} (\in \rep{16}_{2/3})$, $t_R,\, b_R \in \rep{8_s}
(\in \rep{16_s}_{2/3})$ so that all fermions transform under 
$SO(9)\times U(1)_X$ as a $\rep{16  = 8_s + 8_c}$ of $SO(8)$,
both $t_R$ and  $b_R$ belonging to the \rep{(1,3)} of $SU(2)
\times SU(2)$. The Yukawa invariants are in the products 
$\rep{8_V \times 8_s \times 8_c} \in \rep{9 \times 16 \times 16}$. 
\end{enumerate} 

The latter seems more attractive, in particular because it is
naturally consistent with gauge-Higgs unification scenarios. 
In both cases, the fermions couple only to the Higgs doublet with 
opposite $U(1) \subset SO(8)$ charge, so that the other one is a so-called 
inert Higgs\footnote{For a recent discussion of inert Higgs phenomenology 
see \cite{Mrazek:2011iu}.}. Therefore the FCNC effects are suppressed 
because each fermion couples to only one and the same Higgs doublet 
and each fermion can be uniquely embedded in $SO(8)\times U(1)$ 
representations. The model satisfies all the items in our fermion shopping 
list at the beginning of Section~\ref{SU5_fermions_embedding}.


\section{Conclusions}

In this paper we have studied some composite Higgs models with extended
scalar sectors. A detailed analysis has been carried out for the cosets 
$SU(5)/SU(4)\times U(1)$,  and $SO(9)/SO(8)$, which deliver two SM doublets 
or, for $SU(5)/SU(4)$, two SM doublets and one singlet.

In both the $SU(5)$ cases,  a contribution to the $T$ parameter is present 
already at tree level unless there is an unexpected alignment of vevs in the 
scalar potential. A survey of various C2HDM's shows that this is naturally expected 
in other examples of this class of models, unless the symmetries defining the coset are large enough
so that any element in the coset is invariant under an $SU(2)$,
so defining the custodial symmetry.
We observe that  two specific cosets, $Sp(6)/Sp(4)\times SU(2)$ and
$SO(9)/SO(8)$ exhibit such a group theoretical property, 
hence no invariants contributing to $T$ can be written. This
ensures  an effective custodial symmetry at tree level independent from the 
scalar potential.

Turning to fermions, we analyzed the embedding of the matter fields in the 
smallest SU(5) and SO(9) representations, paying attention to the 
effects on the $Zb\bar{b}$ vertex and on the presence of 
Higgs-mediated FCNC. In the SU(5) cases, we find that $Zb\bar{b}$ 
is protected only when the SM fermions are embedded in a \rep{10}, 
while Higgs-mediated FCNC are always absent because 
the non linear Higgs has a single coupling to fermions.
The up and down type quarks couple to different linear combinations
of the Higgs doublets, so that the latter are both active. However the $t$ quark
coupling tends to misalign the Higgs vev's and increases the contribution to $T$.
Therefore, these SU(5) models do need  a compensation mechanism to be phenomenologically viable.

In the SO(9) case an additional abelian charge embedded in SO(8) can be used 
to ensure the fermions to be coupled to just one of the doublets, making the 
other one  inert. 
Moreover, when the left handed quark doublet is embedded in the spinorial 
representation, also the $Zb\bar{b}$ vertex is protected.


\section*{Acknowledgements}
We thank J. Serra, M. Redi and A. Wulzer for useful comments. E.B. thanks M. 
Frigerio and R. Contino for useful discussions. 
This work has been partially
supported by the Agence Nationale de la Recherche under contract ANR
2010 BLANC 041301. The work of TSR is supported by EU ITN, contract
"UNILHC" PITN-GA-2009-237920 and the CEA-Eurotalents program.  HS is
supported by a CAPES Foundation (Ministry of Education of Brazil)
Postdoc Fellowship.


\appendix
\section*{Appendices}

\section{Embedding of the Standard Model fermions in $SU(5)$ representations} \label{app:fermrep}


The SM charges, embedded in the first four entries of the $5\times 5$ $SU(5)$, are given by

\begin{equation}
  T_L^i =   1_{2\times 2} \times \tau^i , ~~T_R^3 = \tau^3 \times 1_{2\times 2}
\end{equation}
with $\tau^i$ the usual $SU(2)$ generators. 
In terms of $(T^3_L, T^3_R)$ charges, this choice corresponds to a
$SU(5)$ fundamental decomposed as

\[
{\bf 5} = \begin{pmatrix} (+,+) \cr (-,+) \cr (+,-) \cr (-,-) \cr
  (0,0) \end{pmatrix}.
\]

The $SU(5)$ fundamental decomposes as $\rep{5}=\rep{(2,2)}+ \rep{(1,1)}$ under
$SU(2)_L \times SU(2)_R$.  For $SU(5)/SU(4)$, the two
LH quarks must be embedded in different bidoublets:
\[
 \begin{array}{ccc}
q_L^t \in {\bf (2,2)_{2/3}} & {\rm with} & t_L = (+,-) \\ 
q_L^b \in {\bf (2,2)_{-1/3}} & {\rm with} & b_L = (-,+) \\ 
u_R \in {\bf (1,1)_{2/3}} & & \\ 
d_R \in {\bf (1,1)_{-1/3}} & & \\
 \end{array}
\]
where in the last column we specified the $(T^3_L, T^3_R)$ assignment.

The specific embeddings are
\begin{equation}
 \Psi_{q_L^t}=\left(\begin{array}{c}

             0_{2\times 1}\\
	      q_L\\
	      0\\
            \end{array}\right)_{2/3} ~~
\Psi_{q_L^b}=\left(\begin{array}{c}

	     q_L\\
	      0_{2\times 1}\\
	      0\\
            \end{array}\right)_{-1/3} ~~
\Psi_{t_R}=\left(\begin{array}{c}

	     0_{2\times 1}\\
	      0_{2 \times 1}\\
	      t_R\\
            \end{array}\right)_{2/3} ~~
\Psi_{b_R}=\left(\begin{array}{c}

             0_{2 \times 1}\\
	     0_{2 \times 1}\\
	      b_R\\
            \end{array}\right)_{-1/3} ~~
\end{equation}

Turning to the $SU(5)$ symmetric representation, it corresponds to $\ten =
\rep{(2,2)}+ \rep{(1,3)} + \rep{(3,1)}$ under $SU(2)_L\times
SU(2)_R$. To accommodate hypercharge, the $U(1)_X$ charge has to be
fixed to $X=2/3$, with assignment
\[
 \begin{array}{ccl}
q_L \in {\bf (2,2)_{2/3}} & {\rm with} & t_L = (+,-), ~b_L = (-,-) \\ 
u_R \in {\bf (1,3)_{2/3}} & {\rm with} & t_R = (0,0) \\ 
d_R \in {\bf (1,3)_{2/3}} & {\rm with} & b_R = (0,-1) \\
 \end{array}
\]
The explicit embeddings are
\begin{equation}
 \Psi_{q_L} = \frac{1}{2}\left( 
\begin{array}{c|c}
 0_{4\times 4} & \begin{array}{c} 0_{2\times 1} \\ q_L \end{array} \\
\hline
\begin{array}{cc} 0_{1\times 2} & q_L^T \end{array} & 0\\
\end{array}\right), ~~
\Psi_{t_R} =\frac{t_R}{2}  \left(
\begin{array}{cc|c}
0_{2\times 2} & i \sigma_2 & 0_{2\times 1} \\ 
i \sigma_2 & 0_{2\times 2} & 0_{2\times 1} \\ 
\hline 
0_{1\times 2} & 0_{1\times 2} & 0
\end{array}\right), ~~
\Psi_{d_R} = \frac{b_R }{\sqrt{2}}\left(
\begin{array}{cc|c}
 0_{2\times 2} & 0_{2\times 2} & 0_{2\times 1} \\ 
 0_{2\times 2} & i \sigma_2 & 0_{2\times 1} \\ \hline
 0_{1\times 2} & 0_{1\times 2} & 0
\end{array}\right).
\end{equation}


\section{Algebraic analysis of the $T$-parameter} \label{app:D&T}

In this Appendix we give a more detailed analysis of the terms corresponding
to anomalous $T$ parameters. The so-called $D$-terms, $D^A= u^{\dag}T^A u$,
where $T^A$ are the generators of a group G in the (arbitrary) representation 
of $u$, have been studied in the context of supersymmetry. For completeness, 
this Appendix contains a the study of the anomalous $T$ parameter defined
in \refe{M2AB} as $D^A D_A$, based on \cite{Buccella:1982nx}, from where we
borrow the following lemma:

{\it Given u defined by its little group  $H\subset G$, $D_A =u^{\dag}T^Au \neq 0$,
if and only if $T^A \in G/H$ and commutes with the elements of H.}

This lemma has been very useful in the study of symmetry breaking in
supersymmetric theories where the fields are complex, even when they 
transform in real representations if they belong to chiral supermultiplets.
Here it is just a curiosity since the calculations are straightforward. 

To apply it in our case, let us take $G=SU(2)\times SU(2)$ and $u \in \rep{(2,2)}$,
and write it as a vector $u_i = x_i + y_i$, $(i=1,...,4)$ such that the  generators are
real, antisymmetric  and act on u as a 4-rotation, $T^{ij}u_k = \delta^{jk}u_i - 
 \delta^{ik}u_j $. Notice that any $u$ can be rotated in the plane $u_1=u_2 =0 $, 
 with little group H = $U(1)$ generated by the (electric) charge, $T^{12}$. 
 Then $T^{34}$ is the only charge in the coset G/H that commutes with 
 $T^{12}$. Indeed, $u^\dag T^{34} u = i(y_3 x_4 - y_4 x_3)$ is the only 
 non-vanishing component of the $D$-terms and, in this basis, $T^{34}$ is the 
 charge corresponding to $Z$ and the anomalous contribution to the $Z$ mass
 is proportional to $(D^{34})^2$. 

Now consider a critical $u$, such that $x_i $ and $y_i$ are aligned and can 
be rotated to $u_1=u_2 = u_3 =0$. Its little group is the vectorial (or custodial) 
$H=SU(2)$ of rotations in this hyperplane. The coset G/H is given by the three axial 
charges that transform as a triplet under H. Therefore all $D^{ij} = 0$ and 
there is no anomalous term in the $Z$ mass. The conclusion is that the 
complex character of the $u$ coordinates, related to $H_2=U(1)$ in the
cosets defining the Higgs doublet as PNGB, is at the origin of the anomalous
$T$ parameter in these cases, unless one can ensure the two Higgs alignment.

The case where $SU(2)\times SU(2)$ is embedded in both $H_1$ and $H_2$,
is different since now the equivalent of the D-terms $D^A= u^\dag T^Au$, are
matrices. Let us concentrate on the relevant instance of $Sp(6)/Sp(4)\times 
SU(2)$  and choose the embedding of the SM subgroup through the $SU(2)^2 
\in Sp(4)$, hence $SU(2)^3 \in Sp(6)$, so that $u$ decomposes as $\rep{(2,1,2)
+(1,2,2)}$ while the bottom components are invariant. Due to the large symmetry, 
the two Higgs in $u$ can be aligned so that the generic $u$ encoding the two 
Higgs PNGB's has a little group the custodial group defined, with some 
convention, as the sum of the three $SU(2)$ charges, defining the custodial 
$SU(2)_C$. Therefore, the $T$ parameter must be non anomalous.

The two Higgs behave as $\rep{(2,2)}$ if  we choose $SU(2)_L \in $ 
the sum of those in $H_1 = Sp(4)$ and $SU(2)_R$ as the remaining
one in $H_2$, or vice-versa. We recall that  the latter `acts from the right'.
After the gauging of the SM symmetries one immediately finds the
masses of the gauge bosons by rotating the Higgs quaternions in $u$ 
to the element $u^T =  (u_0\sin \beta \; {\bf 1}_2, \, u_0\cos\beta \; {\bf 1}_2, \, 
\cos (h/f) \; {\bf 1}_2)$, $u_0 = h \sin(h/f)$, and $h$ is the usual combination
of the Higgs vev's. The result,
\begin{equation}\label{sp6_mass}
\frac{f^2}{2} \left( g^2  W^+_{\mu}W^{-\mu} + \left( gW^0_{\mu} - g' B_{\mu}
\right)^2 \right) \sin^2 (h/f)\cos^2(h/f) ,
 \end{equation}
where the term in $\sin^4 (h/f)$ comes from the quartic term in the PNGB
lagrangian, is clearly consistent with the custodial symmetry, introducing
a correction factor $\cos^2(h/f)$ with respect to the results in the other
2HDM cosets.


\section{Embedding of the Standard Model charges in $SO(9)$} \label{app:app_gen_O9}
In this appendix we describe the embeddings of the $SO(9)/ SO(8)$
coset (PNGB) $\rep{8_V}=\rep{(2,2)}+\rep{(2,2)}\in SU(2)\times SU(2)
\subset SO(8)$. A systematic way to find them is to follow all  the chains of 
maximal subgroups starting from those of  $SO(8)$ down to those that 
contain $SU(2)\times SU(2)$ and discard all those that lead to the wrong
decomposition for the $\rep{8_V}$ of PNGB. We follow the conventions in 
\cite{Slansky:1981yr} and the results are easily checked form the tables 
therein. We simplify the notations by writing $SU(2)^n$ for the $\times$ 
product  of $n \,\, SU(2)$'s. 

Two of the so-defined chains, namely $SO(8) \rightarrow SU(4)
\times U(1) \rightarrow SU(2)^2 \times U(1)$ and $SO(8) 
\rightarrow Sp(4)\times SU(2) \rightarrow SU(2)^2 \times U(1)$, 
yield the decomposition: 
\begin{eqnarray} \label{8-embeddings}
\rep{8_V}=\rep{(2,2)_{+1}}+\rep{(2,2)_{-1}} \qquad 
\rep{8_c}=\rep{(2,2)_{-1}}+\rep{(2,2)_{+1}} \\ 
\rep{8_s} = \rep{(1,1)_{+2}} +\rep{(1,1)_{-2}} + \rep{(3,1)_0} +\rep{(1,3)_0}
\end{eqnarray}
where the $U(1)$ charges are also displayed for each $SU(2)^2$ 
representation.

Three other chains, namely, SO(8)$\rightarrow$SO(7)$\rightarrow$SU(4)
$\rightarrow$SU(2)$^2$,  SO(8)$\rightarrow$SO(7)$\rightarrow$SU(2)$^3 
\rightarrow$SU(2)$^2$ and SO(8)$\rightarrow$SU(2)$^4
\rightarrow$SU(2)$^2$, have the same SU(2)$^2$ decompositions, but since 
there is no U(1) factor, there is no distinction between the two 
$\rep{(2,2)}$'s or the  two $\rep{(1,1)}$'s. 

Finally, note that besides the quadratic invariant  for each $\rep{8}$, 
there is the invariant $\rep{8_s}\rep{8_V}\rep{8_c}$ that will be useful in 
the analysis of Yukawa couplings.


\section{Analysis of the $\mathbf{Sp(6)/Sp(4) \times SU(2)}$ coset}\label{app:Sp6}
\noindent
Sp(6) is the group of unitary transformations such that 
$\Omega \equiv 1_{3} \times \sigma_2$ is invariant , {\it i.e.}, 
$U\Omega U^T = \Omega$. It is useful to work in terms of 
quaternions: $Q= q^\alpha e_\alpha$ , with $e_i = i\sigma_i,\ e_4=1_2$
Notice that:  $Q^{\dagger}Q = \sum q^{i2}\, 1_2$, 
$\sigma_2 Q= Q^*\sigma_2$. Then, the $Sp(6)$ algebra can be 
written as the $3\times 3$ hermitian quaternion matrices with elements 
$iQ_{ij}$ such that $iQ^4_{ii}=0$. The diagonal elements  $iQ_{ii},\, 
(i=1,2,3)$ are then associated with the SU(2)$^3$ sub algebra.  
Adding only $iQ_{12}$ it is upgraded to Sp(4)$\times$SU(2), 
while $iQ_{13}$and $iQ_{23}$ define the coset
Sp(6)/Sp(4)$\times$SU(2), transforming as a \rep{(4,2)} of Sp(4)$\times$SU(2). 
In particular, the associated PNGB matrix $\Pi$ can be written as

\begin{equation}
 \Pi = 
\left( 
\begin{array}{c|c}
        \mathbf{0}_{4} & \begin{array}{c}
               -i\mathbf{H}_1 \\ -i\mathbf{H}_2
              \end{array}\\ \hline
	\begin{array}{cc}
	i\mathbf{H}_1^\dagger & i\mathbf{H}_2^\dagger\\
       \end{array} & \mathbf{0}_{2}\\
\end{array}
\right)
\end{equation}
where the $\mathbf{H}_i \equiv h_i^\alpha e_\alpha $ are the two Higgs doublets reshuffled into the familiar
quaternion form. 

We then define the vector with quaternion components $u_i, ,\ (i=1,2,3)$ 
such that:
\begin{equation}\label{uquaternion}
u  = e^{i\Pi (\mathbf{H}_i)} u(0)  = \frac{\sin(h/f)}{h} \left( 
\begin{array}{c} 
\mathbf{H}_1   \\ \mathbf{H}_2 \\ \cot(h/f) \,\mathbf{1} 
\end{array} \right) \qquad   
u(0)= \left( \begin{array}{c} \mathbf{0}_2 \\ \mathbf{0}_2\\ \mathbf{1}_2  \end{array}
\right)
\end{equation}
\noindent
where $h^2 = \mathrm{tr} \sum_{i=1,...,4} \left(h_1^{i\,2} + h_2^{i\,2}\right)$ 
so that it transforms as defined in Section~\ref{metricnote}, namely,  
$u \rightarrow guh_2^{\dag}(g,\mathbf{H}_i)$ with $g \in G$ and $h_2
\in H_2$. Therefore $u$ can be also viewed as two orthogonal 6-vectors 
$u_{\alpha},\ \alpha=1,2$, each one transforming as the \rep{6} of 
$Sp(6)$, with $u_2 = i \Omega u_1^*$. Notice that, as quaternions, 
 $\Omega u=u^*\sigma_2$.

Another possible parametrization for the breaking Sp(6)$\rightarrow$Sp(4)$\times$SU(2)
is given by the vacuum
\begin{equation}
\Sigma_0 = \left(\begin{array}{ccc} {\bf 0}_2 & & \\ 
 & {\bf 0}_2 & \\ 
 & & \sigma_2 \\
\end{array} \right)\; ,
\end{equation}
\noindent 
in an antisymmetric \rep{14} of Sp(6). Then the n.l.r. can be written in 
terms of the PNBG as $\Sigma = U\Sigma_0  U^{T} = u\sigma_2u^T$.

Following the procedure described in Sections~\ref{SU5_coset} and~\ref{metricnote} to combine 
the invariants to have equivalence 
with the CCWZ construction, the kinetic Lagrangian results

\begin{equation}\label{eq:Sp6_metric}
{\cal L} =\frac{f^2}{4} \left( \partial^\mu u^\dag \partial_\mu u - \partial^\mu u^\dag u u^\dag \partial_\mu u \right)\;,
\end{equation}
so that a tree-level contribution to $T$ is expected from the second term.

Using the $SU(2)^3$ symmetry both quaternions $\mathbf{H}_1$ and 
$\mathbf{H}_2$ can be rotated and aligned along  their components $e_4={\bf 1}_2$,
which is invariant under a  (diagonal) $SU(2)$, the custodial symmetry.
The effectiveness of custodial symmetry can also be explicitly checked 
substituting the ordinary derivatives in \ref{eq:Sp6_metric} with covariant ones: 
$\partial_\mu u \rightarrow \partial_\mu u + i W_\mu^a \left(T^a_L u - u t^a_L \right) + i B_\mu Y u$. 
In the notation of the previous sections, we are specializing to the case in which 
U(1)$_Y\subset$Sp(4) and SU(2)$_L = H_2$, so that the latter acts on $u$ both from the left and from the 
right. It is immediate to write the covariant derivative for the other case.

Turning to the fermions, the smaller Sp(6) representations decompose under SU(2)$\times$Sp(4) as
\begin{eqnarray}
 \rep{14} &=& \rep{(2,4)} + \rep{(1,5)} + \rep{(1,1)} \nonumber\\
 \rep{21} &=& \rep{(2,4)} + \rep{(1,10)} + \rep{(3,1)}
\end{eqnarray}
Since there are several ways to embed the SM fermions, a severe
problem of Flavor Changing Neutral Currents is present.


\end{document}